\documentclass{ws-fnl}
\usepackage{cite}
\usepackage{amsfonts,amsmath,amssymb,graphicx}
\usepackage{amsfonts}
\usepackage{amsmath}
\usepackage{amssymb}
\usepackage{graphicx}
\newcommand{\f}{\frac}
\newcommand{\T}{\mathrm{t}}

\begin{document}


\catchline{}{}{}{}{}

\title{Workshop on noisy many body systems--A cold-atom ratchet interpolating between classical and quantum dynamics}%

\author{\footnotesize R. K. SHRESTHA$^\dagger$, W. K. LAM, J. NI, and G. S. SUMMY}
\address{Department of Physics, Oklahoma State University, \\
Stillwater, Oklahoma 74078-3072, USA\\
$^\dagger$rajendra.shrestha@okstate.edu}

\maketitle
%

\begin{abstract}
We use an atomic ratchet realized by applying short pulses of an optical standing-wave to a Bose-Einstein condensate to study the crossover between classical and quantum dynamics. The signature of the ratchet is the existence of a directed current of atoms, even though there is an absence of a net bias force. Provided that the pulse period is close to one of the resonances of the system, the ratchet behavior can be understood using a classical like theory which depends on a single variable containing many of the experimental parameters. Here we show that this theory is valid in both the true classical limit, when the pulse period is close to zero, as well as regimes when this period is close to other resonances where the usual scaled Planck's constant is non-zero. By smoothly changing the pulse period between these resonances we demonstrate how it is possible to tune the ratchet between quantum and classical types of behavior.
\end{abstract}

\section{Introduction}
Understanding the nature of the crossover between classical and quantum behavior is one of the most important unresolved problems in physics. One place where the stark difference between the two paradigms becomes very clear is in classical non-linear systems. Classically, such a system can exhibit "chaos" in which it is effectively impossible to predict its long term evolution, while in contrast because of the linearity of the Schrodinger equation, an equivalent quantum mechanical systems is completely deterministic. One of the systems of choice for studying this behavior is the so-called delta-kicked rotor, typically realized with a sample of cold or ultra-cold atoms kicked by short pulses of an optical standing wave. This is the atom optics quantum kicked rotor (AOQKR) \cite{aoqr}. While theory and experiment has been successful in elucidating some features of this and similar systems, there are still many aspects that remain to be discovered.
The AOQKR has been  one of the paradigmatic models for studies of experimental quantum chaos and has revealed a wide variety of interesting effects including: dynamical localization \cite{localize}, quantum resonances (QRs) \cite{localize,ryu,fm}, quantum accelerator modes  \cite{fgr,gazal}, and quantum ratchets \cite{monterio,Ratchet,Ratchetp,racht,Ratcheta,Ratchets,Ratchetab,rachtheory,rocking}. The latter are quantum mechanical systems that display directed motion of particles  in the absence of unbalanced forces. They are of considerable interest because classical ratchets are the  underlying  mechanism for some  biological motors and nanoscale devices \cite{racht}.
Recent theoretical \cite{Ratcheta,Ratchetab} and experimental \cite{Ratchet} studies have demonstrated that a controllable ratchet current arises in  kicked atom systems at QR. A QR occurs when the kicking period is commensurate with the natural periods of the rotor and is characterized by a quadratic growth  of the kinetic energy with time. Ratchet behavior away from resonance was investigated in a recent theoretical paper \cite{njp}. In that work, the authors developed a classical-like ratchet theory and  proposed the existence of a one-parameter scaling law that could be used to predict  the ratchet current for a wide variety of parameters. In addition an inversion of the momentum current is possible for some sets of scaling variables. This was experimentally verified in Ref. \cite{raj}.

In this paper, we report on the ratchet current behavior in  the  true classical  and also at the so-called $\varepsilon-$classical regimes, verifying that in both cases, it behaves in essentially the same way, exhibiting an inversion (negative current) for certain families of experimental parameters.  We demonstrate that this can be explained  by  a scaling law when the ratchet is close to the classical limit or one of the QRs. We also show that our model breaks down as the system is moved into the quantum regime.

\section{Theory}

In order to model the AOQKR we introduce a Hamiltonian which in dimensionless units is given by \cite{fgr,ishan,sadg}:  $\hat{\mathcal{H}}=\frac{\hat{p}^{2}}{2}
+\phi_{d}\cos(\hat{X})\sum_{q=1}^{\T}\delta (t'-q\tau)$, where $\hat{p}$ is the momentum (in units of  $\hbar G$, two photon recoils)
 that an atom of mass $M$ acquires from short, periodic pulses of a standing wave with a grating vector $G=2\pi/\lambda_{G}$
($\lambda_{G}$ is the spatial period of the standing wave). The
momentum of this system is only changed in quanta of $\hbar G$, so we
express $p$ as $p=n+\beta$ where $n$ and $\beta$ are integer and
fractional parts of the momentum respectively and the quasi-momentum, $\beta$, is conserved. Other variables are the position
$\hat{X}$ (in units of $G^{-1}$), the continuous time variable $t'$
(integer units), and the kick number $q$. The pulse period $T$ is
scaled by $T_{1/2}=2\pi M/\hbar G^{2}$ (the half-Talbot time) to
give the scaled pulse period $\tau=2\pi T/T_{1/2}$. The strength of
the kicks is given by $\phi_{d}=\Omega^{2}\Delta t/8\delta_{L}$,
where $\Delta t$ is the pulse length, $\Omega$ is the Rabi
frequency, and $\delta_{L}$ is the detuning of the kicking laser
 from the atomic transition.

We are primarily interested in understanding this system close to resonant values of $\tau$ (i.e. $\tau=2\pi\ell$, with $\ell \geq 0$ integer). Here the scaled pulse period is written as $\tau= 2\pi\ell +\varepsilon$, where $|\varepsilon|\ll1$ which measures the closeness to the resonance, and can be shown to play the role of Planck's constant. This allows us to use the so-called $\varepsilon-$classical theory in which the dynamics can be understood by the classical mapping \cite{fgr,sadg},
\begin{equation}\label{map}
  J_{q+1}=J_q+\tilde{k}\sin(\theta_{q+1}),\hspace{4mm} \theta_{q+1}=\theta_q+J_q,
\end{equation}
where $\tilde{k}=|\varepsilon|\phi_d$ is the scaled kicking strength, $J_q=\varepsilon p_q +\ell\pi +\tau \beta$ is the scaled momentum variable and $\theta_q= X \mod(2\pi) + \pi[1-\text{sign}(\varepsilon)]/2 $ is the scaled position exploiting the spatial periodicity of the kick potential. In order to create a ratchet from the above Hamiltonian
 it was shown in \cite{Ratchet} that a superposition of two
plane waves should be used for the initial state. We start with a superposition of two momentum states $|\psi_0\rangle=\f{1}{\sqrt{2}}\left[|0 \hbar G\rangle +e^{i\gamma}|1 \hbar G\rangle\right]$, or equivalently a rotor state $\f{1}{\sqrt{4\pi}}[1+e^{i(\theta+\gamma)}]$. This leads to the position space probability distribution function  $P(\theta)=|\psi(\theta)|^2=\f{1}{2\pi}[1+\cos(\theta+\gamma)]$.  Here $\gamma$  is an additional phase used to  account for the possibility that the initial spatial atomic distribution is shifted in position relative to the applied periodic potential. Although the distribution $P(\theta)$ is quantum in origin, in what follows it will be interpreted as a classical probability.
\begin{figure}[th]
\includegraphics[width=12 cm]{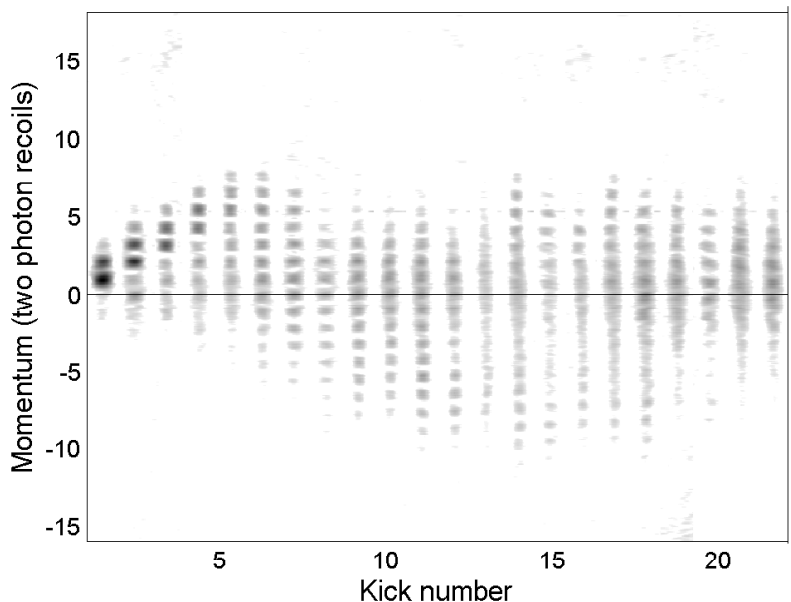}
\caption{Experimental momentum distributions after exposing a BEC to short pulses of an off-resonant standing wave of light. The momentum distributions are shown as a function of  kick number ($\ell=1$, $|\varepsilon| = 0.18$,  $\phi_d= 1.8$ and $\gamma= -\pi/2$.). Each momentum distribution was captured in a separate time-of-flight experiment.}\label{rawplot}
\end{figure}

 One of the first applications of the $\varepsilon-$classical theory to the kicked rotor system showed the existence of a one-parameter scaling law for the mean energy \cite{WimNL}. This was experimentally verified  in the vicinity of the first and second  quantum resonances ($\ell=1$ and $\ell = 2$) in Ref. \cite{mescaling}. It was found that the scaled energy could be written as $\f{E}{\phi_d^2 q}= 1-\Phi_0(x) + \f{4}{\pi x} G(x)$  where $x=\sqrt{\phi_d |\varepsilon|}$ $q$ is a scaling variable and $\Phi_0(x)$ and $G(x)$ are closed form functions of $x$. Recently, the application of such a  scaling law for the ratchet current using the same scaling parameter $x$ was studies \cite{njp,raj}. This work showed the existence of an inversion of momentum current  at some values of the scaling variable (i.e. certain  families of real parameters).

 The theory of this behavior can be derived as follows. First, in the pendulum approximation \cite{pendulum}, the motion of the kicked rotor in continuous time is described by the scaled Hamiltonian $H'\approx (J')^2/2 + |\varepsilon|\phi_d \cos(\theta)$. Here $J'=J/(\sqrt{\phi_d |\varepsilon|})$ is a scaled momentum variable. Near the quantum resonance, using the  position space probability distribution function $P(\theta)$,
 one can calculate $\langle J'-J_0'\rangle=\int_{- \pi}^{\pi}d\theta_0 P(\theta_0)(J'-J_0')$.  For $|\varepsilon| \lesssim 1$,  Eq. (\ref{map}) gives a phase space dominated by a pendulum-like resonance island of extension $4 \sqrt{\tilde k} \gg |\varepsilon|$ \cite{WimNL}. Hence $p=0$ and $p=1$ essentially contribute in the same way giving $J_0'=0$ so that  the map  in Eq. (\ref{map}) is
 $J'_{q+1}= \sqrt{\tilde{k}} \sum_{q=0}^{\T-1}\sin(\theta_{q+1})$.  With the scaling variable $x$, the average scaled momentum becomes \begin{equation} \langle J'-J_0'\rangle= \int_{-\pi}^{\pi}\text{d}\theta_0P(\theta_0)(J'-J_0')=-  \sin\gamma F(x),\end{equation} where

\begin{equation}\label{anas}
  F(x)= \f{1}{2\pi}\int_{-\pi}^{\pi} \sin\theta_0 J'(\theta_0, J_0'=0,x)d\theta_0.
\end{equation}
Thus the mean momentum (units of $\hbar G$) expressed  in terms of the scaled variables is
\begin{align}\nonumber\label{analytic}
  \langle p\rangle=\sqrt\f{\phi_d}{|\varepsilon|}\langle J'-J_0'\rangle &= -\f{\phi_d q \sin\gamma}{x}F(x)\\  \f{ \langle p\rangle}{{-\phi_d q \sin\gamma}}&= \f{F(x)}{x}
\end{align}
 where $F(x)$ can be computed from the above pendulum approximation \cite{njp}.

\section{Experiments}

The experimental results presented here were obtained using the apparatus described in Ref. \cite{raj}. In brief, a  BEC of  about 40000 $^{87}$Rb atoms
in the $5S_{1/2}$, $ F=1$ hyperfine ground state was created in an optical trap \cite{chapman1}. Shortly after being released from the trap, the condensate experienced the potential from a pulsed  horizontal optical standing wave formed by two laser beams of wavelength
 $\lambda=$ 780 nm,  detuned $6.8 $GHz to the red of the atomic transition.   Each beam was aligned at $53^{\text{\textrm{o }}}$ to the vertical to give standing wave wavelength of $\lambda_G=\lambda/(2\sin53^o)$. With these parameters the primary QR (half-Talbot time \cite {ryu,lepers,talbot}) occurred at multiples of $51.5 \pm 0.05$ $\mu $s. Each laser beam was  passed through an acousto-optic modulator  driven by an arbitrary waveform generator. This enabled  control of the phase,  intensity, and pulse length as well as the relative frequency between the kicking beams. Adding two counterpropagating waves differing in frequency by $\Delta f$ results in a standing wave that moves with a velocity $v=2\pi\Delta f/G$. The initial momentum or quasi-momentum $\beta$ of the BEC relative to the standing wave is proportional to $v$, so that by changing  $\Delta f$ the value of $\beta$ could be systematically controlled. The kicking pulse length was fixed at  1.54 $\mu$s, so   we varied the intensity rather than the pulse length to change the kicking strength  $\phi_d$.  This was done by adjusting the amplitudes of the RF waveforms driving the kicking
pulses, ensuring that the experiments were performed  in the Raman-Nath regime (the distance an atom
travels during the pulse is much smaller than the period of the potential).
\begin{figure}[th]
\includegraphics[width=12 cm]{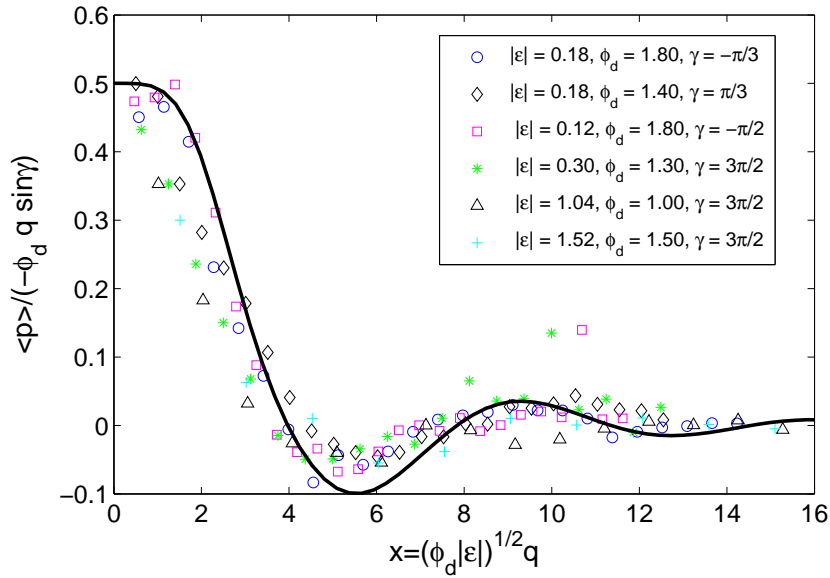}\caption{(Color Online) Scaled mean momentum $\langle p \rangle/(-\phi_d q \sin\gamma)$ as
a function of the scaling variable $x=\sqrt{(\phi_d |\varepsilon|)}q$ for $\ell=1$.  $x$ was varied by scanning over kick number for different combinations of $\phi_d$, $|\varepsilon|$ and $\gamma$.  The solid line  is the function $F(x)/x$ given by Eq. (\ref{analytic}). }\label{scaleplot}
\end{figure}
The initial state for the experiment was prepared as a superposition of two momentum states $|p = 0 \hbar G\rangle $ and $|p = 1 \hbar G\rangle $ by applying a long ($\Delta t=38.6 \mu $s) and very weak standing wave pulse (Bragg pulse).   By using a  pulse of suitable  strength, an equal superposition of the two aforementioned atomic states was created ($\pi/2$ pulse).  The Bragg pulse was immediately followed by the kicking pulses in which  a relative phase  of $\gamma$ between the beams was applied. This phase was experimentally controlled by adjusting the phase difference between the RF waveforms driving the two AOMs. Finally the  kicked atoms were absorption imaged after $9$ ms using a time-of-flight measurement technique to yield momentum distributions like those seen in Fig. \ref{rawplot}. A detailed examination of this data shows that the momentum distributions initially tend strongly towards positive momenta,
followed by  current reversal regions around 15 kicks where the distributions tend  negative. This is  evidence of the current inversion predicted by Eq. (\ref{analytic}).
\begin{figure}[th]
\centering{\includegraphics[width=12.0cm]{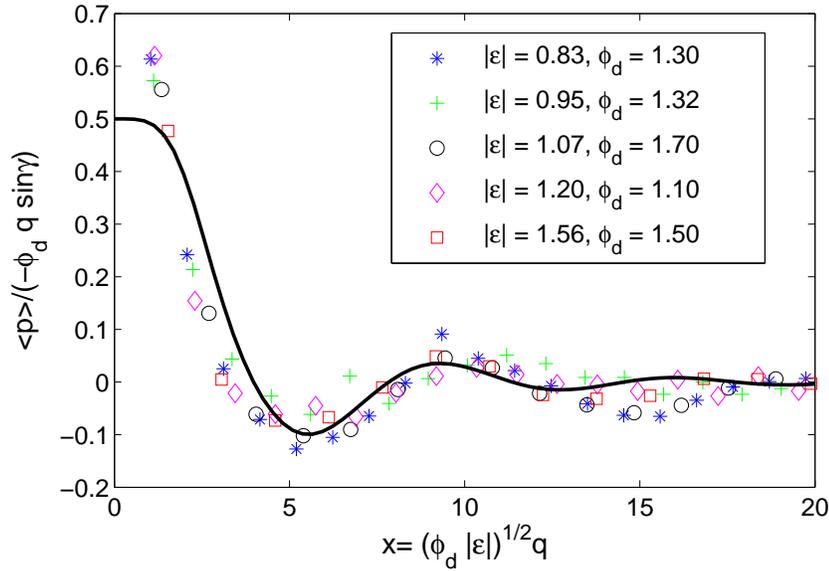}\caption{(Color Online) Scaled mean momentum $\langle p \rangle/(-\phi_d q \sin\gamma)$ as
a function of the scaling variable $x=\sqrt{(\phi_d |\varepsilon|)}q$ for $\ell=0$.  $x$ was varied by scanning over kick number for different combinations of $\phi_d$, $|\varepsilon|$ and $\gamma$. The solid line  is the function $F(x)/x$ given by Eq. (\ref{analytic}).}}\label{scaleplot0}
\end{figure}

\section{Results and Discussion}

The experiments were performed using values of $\tau$ close to the quantum resonance at $\ell=1$ and for $\ell=0$ ($\tau=0$). Since $\tau$ plays a role of an effective Planck's constant,  $\tau\rightarrow0$  is the true classical limit \cite{sadg}.  The measurements involved the determination of the  mean momentum of the kicked BECs  for various combinations of the parameters $q$, $\phi_d$, $\varepsilon$ and $\gamma$. The measured momentum was then scaled by $- \phi_d q \sin\gamma$ and plotted as a function of the scaling variable $x$ for $\ell=1$ (Fig. \ref{scaleplot}) and for $\ell=0$  (Fig. \ref{scaleplot0}). In both figures,  $x$ was changed by varying kick number, $q$ and the solid line  is a plot of the function $\f{F(x)}{x}$ given by Eq. (\ref{analytic}). It can be seen that  the experimental results are in good agreement with the theory for many different combinations of parameters. An exception to this  is seen in Fig. \ref{scaleplot0} at high $x$ values where the experimental data shows the minimum current at higher value of $x (\approx 15)$ different from the value predicted by theory ($x\approx13$). We postulate that this could be due to the systematic errors on the experiment mainly the lack of precision in the measurement of the  kick strength and  the change in kick strength over the longer period of time as the atoms fall through the finite size of the kicking laser beam.  In both cases there is a regime over  $x$  where an inversion of the ratchet current takes place, with a maximum inversion  at $x \approx 5.6$.
\begin{figure}[th]
\center{\includegraphics[width=11 cm]{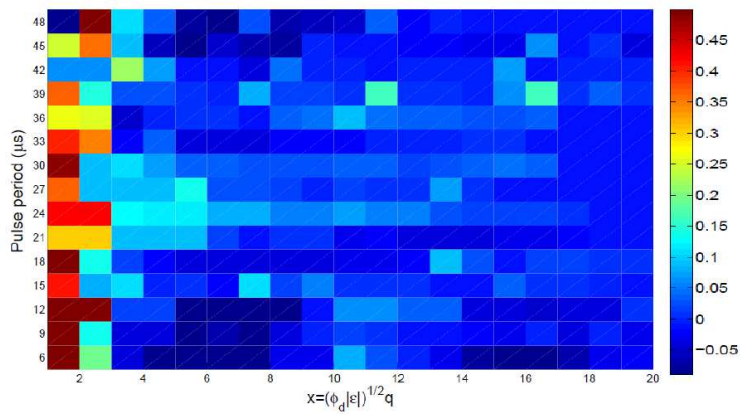}}
\caption{(Color Online) False color plot of the mean momentum current as a function of scaling variable $x$  (the $x-$axis) and pulse period (the $y-$axis). The color scale is the magnitude and direction of the  mean momentum. The deep blue color represents the lowest  value of mean momentum (negative in this case) showing the inversion region. Note that there is a momentum current inversion close to true classical and $\varepsilon-$classical (bottom and top on the $y-$axis respectively) regimes which disappears if one goes away from either (towards the region between $\ell=0$ and $\ell=1$). }\label{pcolor}
\end{figure}
Interestingly this reversal of the ratchet takes place without altering any of the centers of symmetry of the system.  Even though the $\varepsilon-$classical theory assumes $|\varepsilon|$ is small, the experimental results show that it remains valid for higher values of $|\varepsilon|$ as well. This is valid in the true classical regime near $\ell=0$ and in the region where the $\varepsilon-$classical formalism is needed around $\ell=1$. In fact the window of valid $|\varepsilon|$  depends  on the kick number \cite{WimNL}, being rather large for small $q \lesssim 10-15$. This is expected from a Heisenberg/Fourier argument \cite{sadg,mescaling,masa}.

Since the time offset from QR   effectively defines a new Planck constant \cite{fgr,WimNL}, we can easily switch from the classical  to the quantum regime by a simple change of the pulse period \cite{comment}.  Figure \ref{pcolor}  is a   false color plot of scaled mean momentum for the pulse periods starting from close to true classical limit ($\ell=0$) up to the first quantum resonance ($\ell=1$). The data were collected by scanning over the kick number for the different  pulse periods.  The data  presentation is such that the scaling variable $x$ is on the $x-$axis, pulse periods are on the $y-$axis and the  mean momentum is plotted on the color axis. The deep blue color represents negative scaled mean momentum. It can be clearly seen that, when the pulse periods are closer to the classical limit, $\tau=0$ (bottom of the $y-$axis) and to the first quantum resonance $\tau=2\pi$ i.e. $T=51.5\mu$s (top of the $y-$axis), we see the inversion of the momentum current. However the inversion becomes faint and disappears far away from either end (in the middle). This is presumably a result of the system entering the true quantum regime where the classical ratchet theory is no longer applicable. Also visible near $\tau=0$ is a second region of inversion around $x=15$. This is the same as the second inversion seen in Fig. \ref{scaleplot0}. We believe that the second inversion region is not visible near $\ell=1$ due to issues caused by dephasing from vibrations and residual spontaneous emission.

\section{Conclusion}
The experimental results we have presented here show the behavior of an atomic ratchet in the true classical and so called $\varepsilon-$classical regimes. The experiments were carried out by exposing an initial atomic state comprised of
a superposition of two momentum states to a series of standing wave pulses. In both regimes, we measured the scaled mean momentum current as a function of a scaling variable $x$, which contained important pulse parameters such as the offset of the kicking period from resonance, the kick number, and the kick strength. We found that a scaled version of the mean momentum could be described solely by $x$ near $\ell = 0$ and $\ell = 1$ and also verified the existence of momentum current inversions in both regimes.  In other words the true classical and $\varepsilon-$ classical regimes displayed essentially the same behavior. It should be noted however, that in the $\ell = 0$ case, the amplitude of the ratchet current oscillations as a function of $x$ are  more pronounced and are a better match to the theory. We postulate that this is due to the fact that the short time between the pulses near $\ell=0$ provides little opportunity for dephasing effects from vibrations and spontaneous emission to become important. Finally, by continuously changing the pulse period, we were able to observe the breakdown of the $\varepsilon-$classical ratchet theory at pulse periods in-between the two resonances. This breakdown is a signature of the quantum nature of the system.

\section{Acknowledgements}
We  thank Sandro Wimberger for fruitful discussions.


\begin{thebibliography}{99}
\bibitem{comment} At QR there is effectively no evolution between the kicks, making the system formally equivalent to the situation where the pulse period and therefore the scaled Planck's constant are zero.
\end{thebibliography}
\end{document}